\documentclass[reprint,amsmath,amssymb,aps,prb]{revtex4-1}

\usepackage[caption=false]{subfig}
\usepackage{graphicx}
\usepackage{dcolumn}
\usepackage{bm}
\usepackage{xcolor}
\usepackage{braket}

\begin{document}

\title{The Proximity Effect in a Superconductor-Quasicrystal Hybrid Ring}

\author{Gautam Rai}
 \email{gautamra@usc.edu}

\author{Stephan Haas}%

\affiliation{%
Department of Physics and Astronomy, University of Southern California, Los Angeles, California 90089-0484, USA
}%

\author{Anuradha Jagannathan}
\affiliation{
Laboratoire de Physique des Solides, Universit\'{e} Paris-Sud, 91405 Orsay, France
}%

\date{\today}

\begin{abstract}

We compute the real-space profile of the superconducting order parameter (OP) in a hybrid ring that consists of a 1D superconductor connected to a Fibonacci chain using a self-consistent approach. In our study, the strength of the penetration, as measured by the order parameter at the center of the quasicrystal, depends on the structural parameter $\phi$, or phason angle, that characterizes different realizations of the Fibonacci chains of a given length. We show that the penetration strength dependence on $\phi$ reflects properties of the topological edge states of the Fibonacci chain. We show  that the induced superconducting order parameter averaged over all chains has a power law decay as a function of distance from the S-N interface. More interestingly, we show that there are large OP fluctuations for individual chains and that the penetration strength in a finite Fibonacci chain can be significantly $larger$ than in a normal periodic conductor for special values of $\phi$.  

\end{abstract}

\maketitle

	\section{Background}
	
	The proximity effect at a junction between a superconductor (S) and a normal conductor (N) is a phenomenon where the superconducting order parameter is induced in the normal conductor by, loosely speaking, the leakage of Cooper pairs across the interface. The proximity effect can be a sensitive probe of the electronic properties of the latter, for example it has recently been discussed as a possible means to detect Majorana fermions in a suitably engineered hybrid system \cite{fu2008superconducting}. The proximity effect in S-N hybrid systems for graphene monolayers and bilayers are some other examples that display the effects of the unconventional band structures in these materials \cite{komatsu2012superconducting, black2008self, ojeda2009tuning}. We choose here to consider the proximity effect in quasicrystals, whose electronic properties are known to differ greatly from those of periodic solids on the one hand, and from disordered solids on the other \cite{janot1997quasicrystals}. Specifically, we consider a 1D hybrid system consisting of a BCS-type superconductor coupled to a Fibonacci chain. We show that in this case there is a strong proximity effect, with superconducting correlations decaying as a power of the distance from the interface. We will show that the strength of the penetration is related to topological features that are known to be present in the Fibonacci chain \cite{tanese2014fractal, baboux2017measuring}. Inspired by experiments which were carried out for disordered conductors \cite{gueron1996superconducting} we propose a scenario to test these theoretical predictions in a mesoscopic-scale experiment with quasicrystals. 
	
	To set the stage and explain the motivations for this study, first recall some properties of quasicrystals. Although 3D quasicrystalline alloys are conductors, their conduction is non-metallic. Electronic states in quasicrystals are typically neither extended (like Bloch states of periodic metallic crystals) nor exponentially localized (as in disordered crystals). The Fibonacci chain, a one-dimensional quasicrystal, is the best understood with many known results for the single electron spectrum and states \cite{kohmoto1987critical}. We consider the pure hopping Fibonacci chain (FC) model, in which the hopping amplitudes take two possible values $t_A$ and $t_B$, according to a deterministic rule. A single parameter, the ratio $t_A/t_B$, determines the nature of the energy spectrum which is fractal, with gaps everywhere, except for the case $t_A/t_B=0$ when the spectrum consists of three highly degenerate levels, and the periodic case $t_A/t_B=1$ when the spectrum is continuous. For all other values of $t_A/t_B$, the spectrum and all wavefunctions are fractal (see for example a recent discussion in \cite{mace2016fractal}). The gaps of the spectrum can be labeled by integers, according to the gap labeling theorem \cite{bellissard1989spectral}. That these labels correspond to experimentally accessible topological---winding---properties of edge states has been shown in several recent studies \cite{tanese2014fractal, baboux2017measuring}.
	
	We will show that the penetration of the superconducting order parameter into the bulk of the quasicrystal can be very significantly enhanced due to topological edge states. Edge states can, in principle, be created and manipulated by means of a tuning parameter $\phi$, analogous to the phase factor which appears in the potential energy term in the Harper model \cite{aubry1980analyticity, harper1955single}. This parameter, which we will define later, creates a succession of geometric changes (termed phason flips) in the chain, with a concomitant change of its band structure. By cycling through values $\phi$ in the interval $[0,2\pi)$, one can create or annihilate edge states, transport them from one end to the other, and measure their winding numbers. The winding numbers are topological indices characteristic of the gaps of the spectrum, via the gap labeling theorem \cite{bellissard1989spectral}. The different realizations of the Fibonacci chain as a function of $\phi$ have been experimentally realized in polaritonic system \cite{tanese2014fractal, baboux2017measuring}, and the topological indices were directly observed for the first time by imaging states in these chains. In the next section we summarize some of the properties of FC needed for our study of the proximity effect.
	
		\begin{figure*}
		\subfloat[\label{ring}]{
\includegraphics[width=0.32\textwidth]{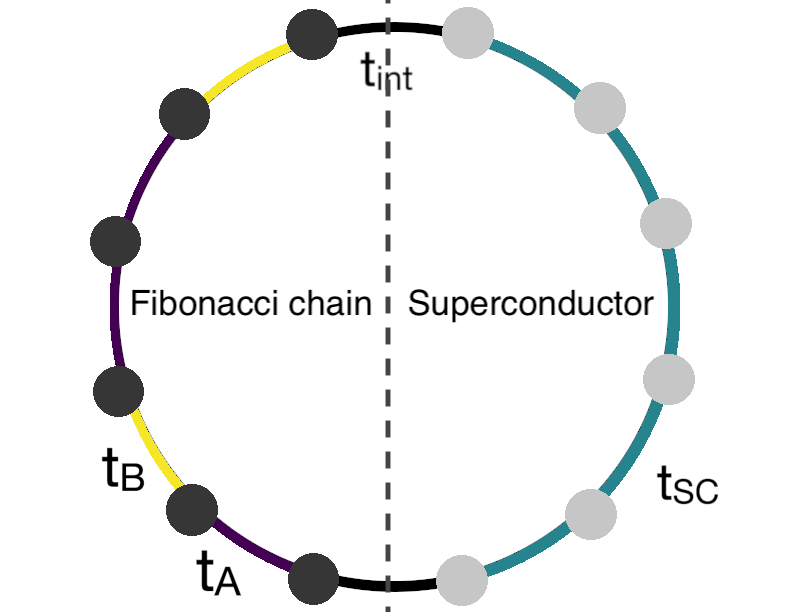}}
\subfloat[\label{chainsbyphi}]{
\includegraphics[width=0.3\textwidth]{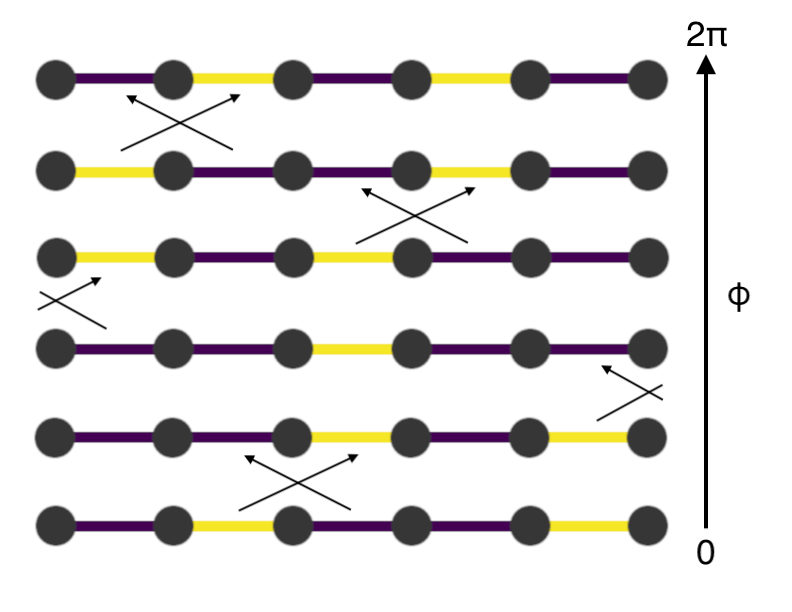}}
\subfloat[\label{idos}]{
\includegraphics[width=0.32\textwidth]{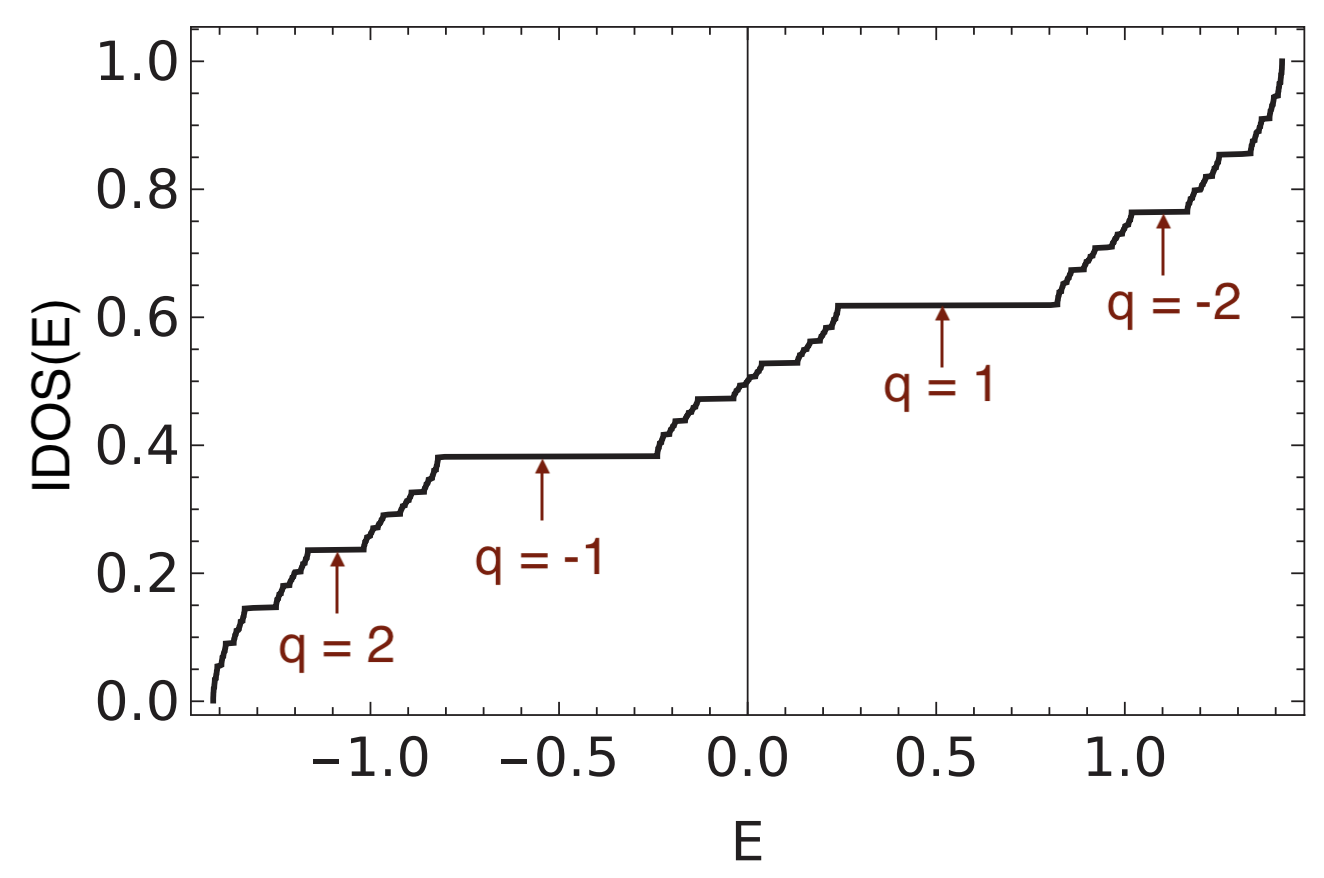}}

\caption{a) A sketch of a quasicrystal-superconductor hybrid chain. On the left is a Fibonacci chain with nearest neighbour hopping terms that take one of two values $t_A, t_B$ modulated according to the Fibonacci chain; on the right is a superconductor with constant hopping $t_{SC}$ and on site attraction $V$ between opposite spins. b) Successive phason flips ($AB \to BA$) in the 6-site Fibonacci chain as $\phi$ is varied. a) The normalized integrated density of states for the Fibonacci chain. The four largest gaps are marked with their $q$ labels. Figure taken from \cite{mace2017critical} with modifications. }
\end{figure*}
	
\section{The Fibonacci chain}
		The Fibonacci string $S_\infty$ is obtained as the infinite limit of the sequence of strings on the two-letter alphabet $\{A,B\}$ defined by $S_0 = A$, $S_1 = AB$, $S_{n+1} = Concatenate(S_{n},S_{n-1})$\footnote{\textit{Concatenate} is a term borrowed from formal language theory and appears in the literature of quasicrystals. In this context, \textit{concatenating} two strings refers to placing them one after the other to form a longer string. e.g. $Concatenate(AB,A) = ABA$}. The $n^\textrm{th}$ sequence $S_n$ has length $F_{n+2}$ where $F_n$ is the $n$th Fibonacci number. $S_\infty$ can also be generated via the characteristic function $\chi_j$ which, if we identify $-(+)$ with $A(B)$, defines the $j$th letter in $S_\infty$.
		
\begin{align}
		\chi_j = sign\left[\cos(2\pi j \tau^{-1} + \phi_0) - \cos(\pi \tau^{-1})\right] \label{chij}
\end{align}	
where $\tau = \frac{1 + \sqrt{5}}{2}$, $\phi_0 = \pi \tau^{-1}$, and $j = 1,2,3, \ldots$. 

If $j$ is terminated at a Fibonacci number $F_n$, we recover the Fibonacci approximant $S_{n-2}$ from before. 
Replacing the constant $\phi_0$ by a variable $\varphi \in [0,2\pi)$, one can generate a family of ($F_n+1$) chains of length $F_n$ corresponding to different values of $\varphi$ (see Fig \ref{chainsbyphi}). In the following discussion, we will use the convenient parametrization $\phi = (\varphi - c_L) \mod 2\pi$ where $c_L$ is a length dependent constant\footnote{$c_L =  (N + 1)\pi\tau^{-1}$}.

The pure hopping Hamiltonian for a given FC realization is $H = \sum\limits_{i} -t_i (c^\dagger_{i+1} c_{i} + c^\dagger_i c_{i+1})$. where the hopping amplitudes $t_i$ have the values $t_A$ ($t_B$) if the $i$th letter of the sequence is A(B). Here the spin index is not explicitly written, as the two spin sectors are identical.

The energy spectrum of $FC_\infty$ is known to be a fractal, specifically a Cantor set of Lebesgue measure zero \cite{suto}. There are an infinite number of gaps and they are dense everywhere in the spectrum. The gap labelling theorem tells us that each gap can be labelled by a unique integer $q$ \cite{bellissard1989spectral}. This is best seen via the integrated density of states in Fig \ref{idos}.

\begin{equation}
IDOS(\epsilon)\big|_{\epsilon \in \textrm{gap with label q}} = q\tau^{-1}\bmod 1 = q\tau^{-1} + p
\end{equation}

where $\tau$ is the golden ratio as before. It suffices to specify one single integer, $q$, as then $p$ is fixed by the condition that the IDOS lie between $0$ and $1$.

The spectrum of the Fibonacci chain can be constructed recursively using the RG scheme of Niu and Nori \cite{niu1990spectral} where one considers a sequence of periodic chains built from approximants to the Fibonacci chain converging to $FC_\infty$. From this, one recognizes a hierarchy in the set of gaps: smaller gaps correspond to larger $|q|$s and first arise in longer approximants \cite{piechon1995analytical, mace2017gap}.

The $q$ labels show up again in the behaviour of edge states in the open Fibonacci chain (or any other geometry that hosts edge states). The energy of an edge state living in a gap with label $q$ will traverse the gap from one end to the other $q$ times as $\phi$ is varied from $0$ to $2\pi$ (see Fig \ref{vals}). In \cite{tanese2014fractal, baboux2017measuring}, this behaviour has been observed in the photoemission spectrum of a polaritonic Fibonacci quasicrystal.

	\section{The SC-FC hybrid ring}

Consider now a hybrid system that consists of a Fibonacci chain attached to a 1d superconducting chain at zero temperature. The Fibonacci chain, consisting of $N_n$ sites, is described by a 1d tight binding model as in the last section. The superconductor, consisting of $N_s$ sites, will be described by a Hubbard model with attractive onsite interactions. The Hamiltonian for the full system is the sum of three terms  $H = H_{FC} + H_{SC} + H_{int}$ (standing for the FC, the SC and an interfacial coupling) and it takes the following form: 
\begin{align}
    H =& \sum_{i,\sigma} \left(-t_i (c^\dagger_{i+1,\sigma} c_{i,\sigma} + c^\dagger_{i,\sigma} c_{i+1,\sigma}) + (-\mu_i + U_i)c^\dagger_{i,\sigma} c_{i,\sigma} \right) \nonumber \\
&- V \sum_{i\in SC}\left(  c^\dagger_{i\uparrow} c^\dagger_{i_\downarrow}c_{i\uparrow} c_{i\downarrow} + h.c. \right)
\end{align}

where $i=1, 2, \ldots, N_n+N_s$ runs over all the sites of the system.
The chemical potential $\mu_i=0$ in both the superconductor and the Fibonacci chain, so that particle-hole symmetry holds, and maintained throughout the self consistent loop calculations. The $t_i$ take  values $t_A$ or $t_B$ in the Fibonacci chain, and an a priori constant value within the superconductor. To reduce the number of parameters in our calculations, we fix the overall energy scale given by $t_B$ to the value $1$. We also fix the hopping amplitudes in the superconductor and at the interface to $t_i=1$, i.e. we assume perfect transmission at the interface. This leaves us with a single hopping parameter $t_A$ to vary in the problem. To date, it appears that superconductivity in quasicrystals is rather uncommon, and corresponds to extremely low critical temperatures as seen in the recent example reported by
K. Kamiya and others\cite{kamiya2018discovery}. This suggests that interaction effects are weak in these materials. Thus, in our model, so that there is no intrinsic superconductivity in the quasicrystal, we set the on-site attractive potential to $0$ in the Fibonacci chain and constant $V$ in the superconductor. $V$ is taken to be the same order of magnitude as the hopping terms so that relevant physics is accessible at computationally accessible system sizes. The results shown in the next section correspond to $V = 0.8$. With these choices, the weak hopping $t_A$ and the phase $\phi$ are the only remaining variable parameters.

The Hamiltonian $H$ is now linearized using the real space Bogoliubov mean field\footnote{While the base problem is 1D, in an experiment, the hybrid ring will be embedded in a larger 3-dimensional system. Therefore, a mean-field treatment should give qualitatively correct results.} approach which is standard for problems lacking translation invariance as in disordered superconductors \cite{ghosal2001inhomogeneous,altomare2013one}. Introducing the spatially dependent mean field superconducting order parameter $\Delta_i = \braket{c_{i\downarrow}c_{i\uparrow}}$, one has

\begin{align}
H_{mf} =& \sum_{i,\sigma} \left(-t_i (c^\dagger_{i+1} c_i + c^\dagger_i c_{i+1}) + (-\mu_i + U_i)c^\dagger_{i} c_i \right) \nonumber\\
&- V \sum_{i\in\text{SC}}\left(  \Delta_i c^\dagger_{i\uparrow} c^\dagger_{i_\downarrow}\right) + h.c.
\end{align}

The order parameter $\Delta_i$ and the on-site potential $U_i$ are self-consistently computed using the Bogoliubov-de Gennes self consistency equations which take the following form at zero temperature:


\begin{align}
\Delta_i =& V\sum_n v^\dagger_{n,i} u_{n,i}\,\textrm{sign}(E_n)\\
U^{sc}_i =& -V\left(\sum_{E_n<0} |u_{n,i}|^2 + \sum_{E_n>0}|v^\dagger_{n,i}|^2 \right)
\end{align}	\label{BdG}

where the sum runs over the eigenstates of the Hamiltonian labelled by $n$. $u_{n,i}$ and $v_{n,i}$ are respectively the $n^\textrm{th}$ particle and hole eigenstates of the Hamiltonian with energy $E_n$. We initially set $\Delta_i$ to be zero in the Fibonacci chain and constant in the superconductor, and $U_i$ to be zero everywhere. Then we diagonalize the system, compute the new order parameter and on-site potential and iterate until convergence.

\section{Results}

\begin{figure}
\includegraphics[width= 0.45\textwidth]{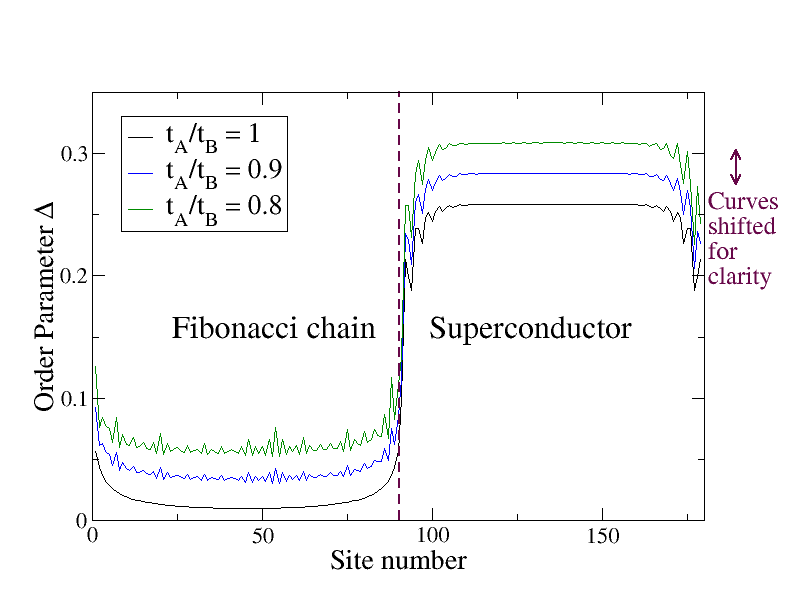}
\caption{\label{OPofx}Self consistently computed order parameter in real space for various $t_A/t_B$. The 90-site quasicrystal is on the left, and the 90-site superconductor is on the right. The blue(green) curve has been shifted up by 0.025(0.05) for clarity.} 
\end{figure}

\begin{figure}
\includegraphics[width= 0.45\textwidth]{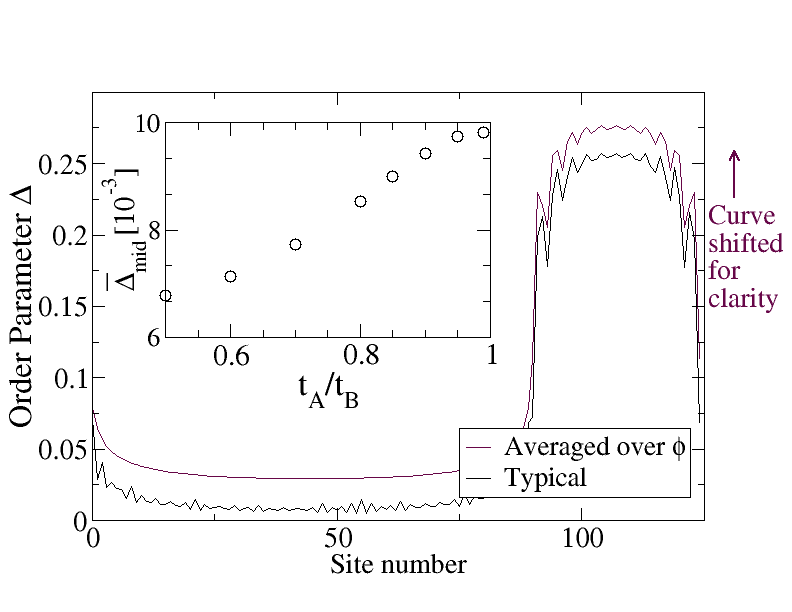}
\caption{\label{typical-vs-average}The order parameter profile averaged over $\phi$ compared with a typical OP profile for $t_A/t_B = 0.9$. The 90-site quasicrystal is on the left, and the 35-site superconductor is on the right. The red curve has been shifted up for clarity. Inset: The average penetration strength, as measured by the OP amplitude at the center of the chain in the $\phi$-averaged profile $\bar{\Delta}_{mid}$, as a function of the modulation strength} 
\end{figure}

In Fig \ref{OPofx}, we show the real-space profile of the order parameter (OP) for a representative system: a $90$-site superconductor connected to a $90$-site quasicrystal for three values of $t_A/t_B$. The left half of the figure shows the order parameter in the quasicrystal, and the right half shows the order parameter in the superconductor. Taking up the superconducting side first, we see that, as expected, the order parameter has a constant value in the bulk, while it gets suppressed closer to the interface due to the so-called inverse proximity effect. Superposed on top of the decay in the OP, we see Friedel-type oscillations. Coming now to the normal side, the profile of the OP depends on the hopping ratio: while it is a smooth curve in the homogeneous case ($t_A/t_B = 1$), spatial fluctuations  become stronger as the ratio $t_A/t_B$ decreases, i.e., as the quasiperiodic modulation gets stronger. 


\begin{figure*}\subfloat[\label{ctrbyrho}]{
        \includegraphics[width = 0.32\textwidth]{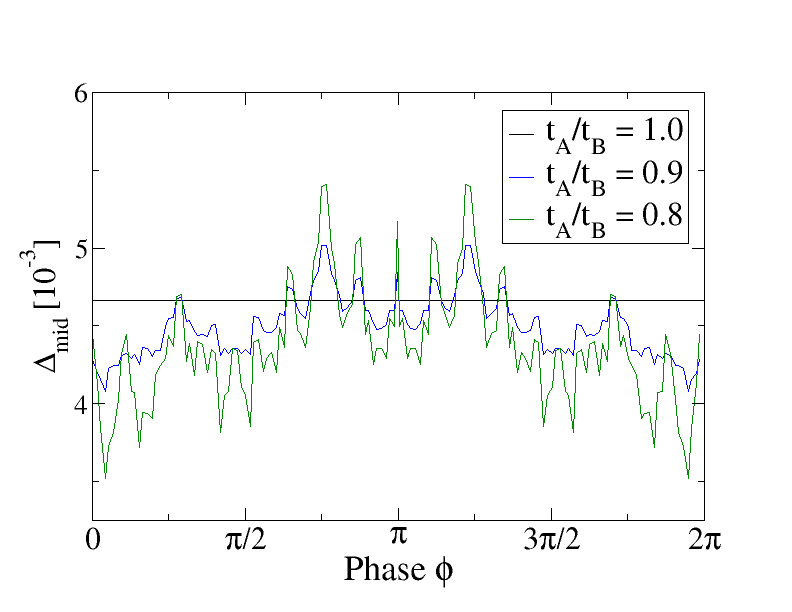}
        }
        \subfloat[\label{ctrbyN}]{
		\includegraphics[width=0.32\textwidth]{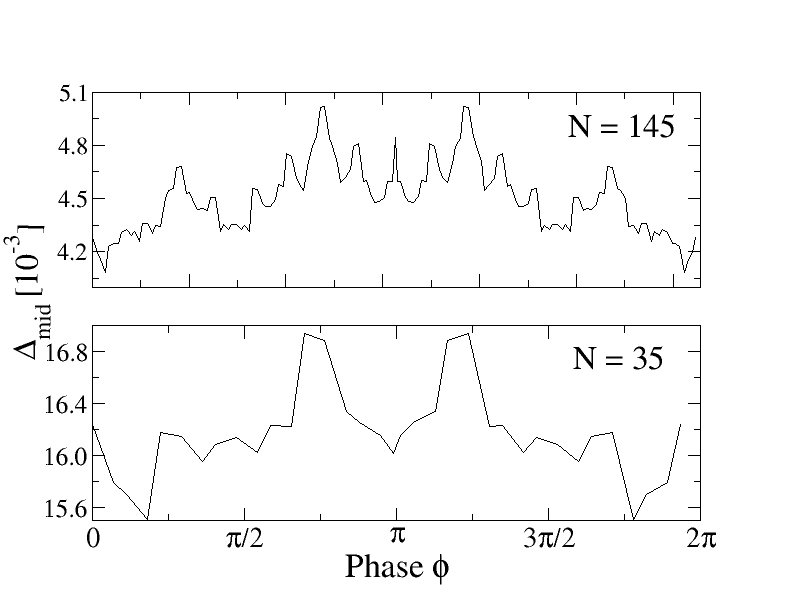}}
		\subfloat[\label{fourier}]
		{\includegraphics[width=0.32\textwidth]{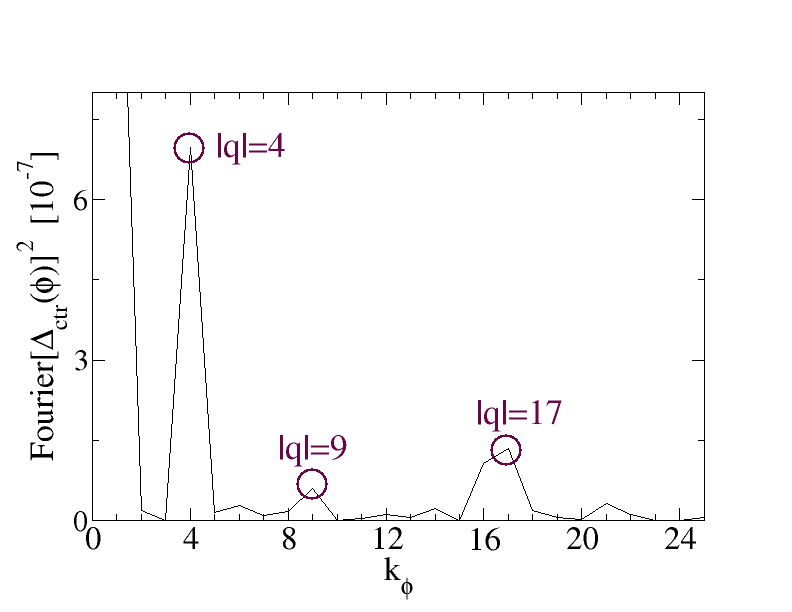}
		}
		\caption{a) \& b) The penetration strength $\Delta_{mid}$ of the order parameter into the Fibonacci chain as a function of the $\phi$ used to generate the chain. In (a), the FC is 145-site long and the different colors represent different modulation strengths. In (b), the modulation is constant $(t_A/t_B = 0.9)$, but the length of the quasicrystal $N$ is varied. c) Power spectrum of the penetration strength (blue curve in Fig \ref{ctrbyrho}) $|\mathcal{F}[\Delta_{mid}(\phi)]|^2$  with the prominent peaks highlighted.}
\end{figure*}

For the quasiperiodic case, $t_A/t_B < 1$, the OP profile is different for each of the $N_n$ different members of the family of FCs of length $N_n$. Recall that each member of this set is characterized by a phase angle $\phi$. We consider first the average behaviour of the OP given by the quantity  $\bar{\Delta}_i = \frac{1}{N_n}\sum\limits_{j=1}^{N_n} \Delta_i^{\phi_j}$ where $\Delta_i^{\phi_j}$ is the order parameter profile of the system generated with the phase $\phi_j$. In Fig \ref{typical-vs-average}, we show a typical order parameter profile for $t_A/t_B=0.9$ compared with the order parameter profile averaged over $\phi$. Note that the averaging process smooths out the order parameter fluctuations in the quasicrystal. To investigate the effect of varying the hopping ratio, we will use the amplitude of the order parameter at the center (in even chains, that of the two central sites) $\Delta_{mid}$ as a measure of the strength of the penetration of the superconducting order into the Fibonacci chain. The inset of Fig \ref{typical-vs-average} shows that the average penetration strength monotonically decreases as $t_A/t_B$ decreases. This is consistent with the intuitive picture according to which wavefunction correlations decay faster, on the average, when the quasiperiodic modulation gets stronger.

Next, we consider the results for individual realizations for different $\phi$ values. Directly fitting the OP curves turns out to be difficult and gives unreliable results due to the strong oscillations. We therefore proceed in two steps: first fitting $\Delta_i$ to a power law given by the expression \eqref{fit}, and then evaluating the fit function at the center of the chain to obtain $\Delta_{mid}$.
\begin{align}
		\Delta^{fit}_i &=  c + b(i^{-\delta} + (L + 1 - i)^{-\delta})\label{fit}
\end{align}

$L$ represents the effective FC chain length, as the location of the interfaces are slightly displaced with respect to the interfaces in the uncoupled problem. The best fits, with the smallest mean square deviation, were found for $L=N_n+2$. We have assumed a symmetric decay with respect to left ($i=1$) and right ($i=L$) edges (the $i$s are appropriately relabelled to account for the shifted interfaces).

Before discussing the results of the fitting procedure, we wish to make a note on the choice of the fit function \eqref{fit}. While edge states in the Fibonacci chain are localized and should decay exponentially, we do not necessarily expect the order parameter to decay exponentially. This is because it is a sum of contributions from all states in the system. The form of each individual contribution is a product of particle and hole wave functions of the form $v^\dagger_{n,i} u_{n,i}$ and they are free to oscillate between positive and negative values in or out of phase with each other as a function of position. We find that the curves we get from our simulation are fit much better by \eqref{fit} than an exponential by every goodness-of-fit metric.

Across the systems we studied, the fit parameter $\delta$, the decay power, lay in the range $0.3-1.0$ depending on the system parameters and the details of the fit. The value of $\delta$ for a given set of system parameters varies significantly depending on how the fit is taken. For instance the parameter set $(N_n = 90$, $N_s = 35$, $t_A = -0.9)$ leads to the fit parameters $( \delta = 0.7330, b = 0.09285 , c = -0.003255)$ if $L$ is taken to be $N_n+2$, and the fit parameters $(\delta = 0.4258, b = 0.06979, c = -0.0181)$ if $L$ is taken to be $N_n$. For this reason, we have chosen to use the fit-independent, appropriately smoothed order parameter amplitude at a given point (the center of the chain) as our measure for the strength of the proximity effect.

\begin{figure}
\includegraphics[width=0.5\textwidth]{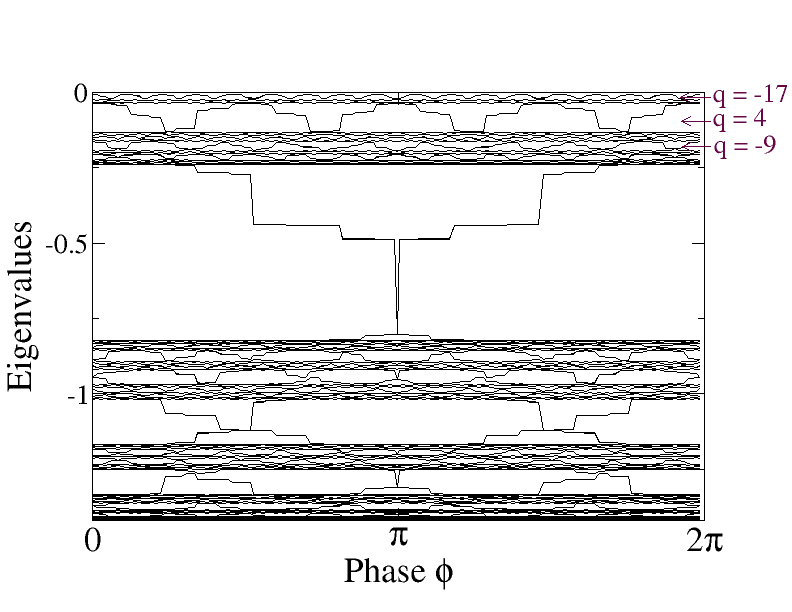}
\caption{\label{vals}The lower half of the spectrum of the open Fibonacci chain of length $145$ as a function of $\phi$ with $t_A/t_B = 0.5$. The gaps closest to the Fermi level are highlighted.}
\end{figure}

    In Fig \ref{ctrbyrho}, we show the results thus obtained for $\Delta_{mid}$ as a function of $\phi$ for the set of FC of length $N_n=145$. The different colors represent different modulation strengths. These curves show that for fixed system size, $\Delta_{mid}$ fluctuates with increasing amplitude as the quasiperiodic modulation increases. In Fig \ref{ctrbyN}, we show this quantity as a function of $\phi$ for FCs of length $35$ and $145$ with modulation $t_A/t_B = 0.9$. A few remarks are in order. Firstly, for particular values of $\phi$, $\Delta_{mid}$ is higher in the quasicrystal than it is in the periodic chain, i.e. for these values of $\phi$ the superconducting order parameter travels further into the quasicrystal than it does into the periodic chain. Secondly, there appear to be some oscillations which conserve the same period as one goes from smaller to longer chains.
	
	To understand the mechanisms which lie behind these observations, we  look at the Fourier components of the curves in Fig \ref{ctrbyN}. Fig \ref{fourier} shows the power spectrum of the blue curve in Fig \ref{ctrbyrho}, ($t_A/t_B = 0.9, N_n = 145$). One sees that there are strong peaks corresponding to periods of 4, 9 and 17 along with smaller peaks not labeled in the plot. The plot in Fourier space suggests a relation between the delta exponent and the edge states in the main gaps which are labeled by winding numbers $q$ as the numbers 4, 9 and 17 are special for the Fibonacci chain. Consider Fig \ref{vals}, which shows how the energy eigenvalues of the 145-site open Fibonacci chain vary with $\phi$\footnote{We have checked, by turning on the interface coupling slowly and directly looking at the wave functions and energy eigenstates of the coupled problem, that the Fibonacci gaps along with their $q$-labels are preserved in the coupled problem.}. The x-axis represents the values of $\phi$, which vary for different members of the family, while the  y-axis corresponds to the set of energy eigenvalues of the chains. One sees clearly how the energy levels of certain states---corresponding to edge states---wind within the gaps and it is easy to check that the number of gap crossings for a given gap is nothing but the index of the gap. The prominent gaps closest to half-filling are expected to contribute the most to the OP -- these, as can be seen in the figure have the labels $q = -17, 4$ and $-9$, precisely the periodicities we saw to be present in the power spectrum of the OP.

\section{Discussion and Conclusion}
We have carried out a self-consistent calculation of the real-space profile of the superconducting order parameter, in a superconductor-quasicrystal hybrid ring. We find that $\Delta_i$ decays as one leaves the S-N interface towards the interior of the quasicrystal, however the decay is not monotonic and has strong fluctuations. The penetration strength, as measured by the order parameter amplitude at the center of the chain, $\Delta_{mid}$, depends on the phase $\phi$ associated with the chain. We find that for some values of $\phi$ the order parameter will travel further into the quasicrystal than into the periodic chain. Moreover the dependence of the penetration strength on $\phi$ has a fractal structure with a hierarchy of smaller peaks surrounding the larger ones. The positions of the peaks correspond directly with the positions of level crossings in the gaps in the spectrum of the corresponding open Fibonacci chain, and the periodicities of $\Delta_{mid}(\phi)$ are exactly the $q$ labels of the gaps closest to the Fermi level. Since the $q$ label is an inherited property of the edge state living in the gap, this indicates a direct correspondence between the penetration strength and the edge state and suggests the possibility of using the proximity effect as a probe for the topological edge states of the Fibonacci chain in particular, and higher dimensional quasicrystals as well as other systems that exhibit topological edge states. Although our method involves a mean-field approximation, in an experiment, the hybrid ring will be embedded in a larger 3D structure, which will serve to stabilize the order parameter and our results should qualitatively agree--there should be large fluctuations in the strength of the induced order parameter based on the quasiperiodic realization and there should be topological edge effects.

We also found that the average values of the induced superconducting order parameter have a smoother spatial profile, and can be fitted to power law decay with a non-universal power law $\delta$. The decay is faster as the quasiperiodic modulation is increased. This is in accordance with the expectation that wavefunctions become less extended as the modulation is increased, leading to rapid decay (on the average) of spatial correlation functions. Such power laws are expected on general grounds, based on the local multifractal properties of wavefunctions in a quasicrystal. Work on theoretical solutions and specific predictions for this model is in progress and will be reported elsewhere. It will be interesting, as well, to investigate whether adding a certain amount of disorder to the quasicrystal results in a longer range of penetration, since disorder smears out the strong quantum interferences present in the perfect quasicrystal. The analogous problem for electronic transport is well-studied in the literature, where it is known, both from experiments and numerical calculations, that adding disorder paradoxically leads to an increase of the conductivity of perfect quasicrystals \cite{mayou1993evidence}.

Threading the hybrid ring with a magnetic flux and studying the structure of the resulting oscillations is another fascinating question reserved for future study.

In conclusion, we suggest that it will be interesting to study the proximity effect and associated topological signatures experimentally. One possibility for experiment consists of inducing superconductivity in quasiperiodic flakes. Monolayers of quasiperiodically organised lead atoms have in fact been successfully deposited on specially chosen surfaces, and a number of other promising systems are possible as described in \cite{ledieu2008self}. If such mesoscopic flakes could be put into contact with larger superconducting lead islands, the resulting superconducting gaps could then be investigated via STM. Although we expect that qualitative features of our results will persist in two dimensions, we plan to extend our approach so as to obtain results specific to 2D systems in future work. 

\begin{acknowledgements}
S.H. acknowledges support from DOE under
Grant No. DE-FG02-05ER46240. G.R. would like to acknowledge a helpful discussion with Lorenzo Campos Venuti.
\end{acknowledgements}

\end{document}